\documentclass[journal, 11pt, onecolumn]{IEEEtran}

\usepackage{cite}
\usepackage{url}

\usepackage{cite}
\usepackage{url}

\usepackage{amssymb}
\usepackage{amsmath,bm}
\usepackage{amssymb}
\usepackage{epsfig}
\usepackage{epsf}
\usepackage{subfigure}
\usepackage{graphicx}
\usepackage{algorithm}
\usepackage{algorithmic}
\usepackage{url}
\usepackage{amsthm}

\usepackage[bottom]{footmisc}

\date{}
\def\BibTeX{{\rm B\kern-.05em{\sc i\kern-.025em b}\kern-.08em
    T\kern-.1667em\lower.7ex\hbox{E}\kern-.125emX}}

\newtheorem{theorem}{Theorem}
\newtheorem{definition}{Definition}

\newtheorem{lemma}{Lemma}
\newtheorem{proposition}{Proposition}
\newtheorem{remark}{Remark}

\newcommand{\Text}[1]{\text{\textnormal{#1}}}

\newenvironment{lemmarep}[1]{\noindent {\bf #1.}\begin{it}}{\end{it}}

\author{Sina Lashgari and Amir Salman Avestimehr 
\thanks{S. Lashgari is with the School of Electrical and Computer Engineering, Cornell University, Ithaca, NY (email: sl2232@cornell.edu); and A. S. Avestimehr is with the EE Department of University of Southern California, Los Angeles, CA 90089 (email: avestimehr@ee.usc.edu). The research of A. S. Avestimehr and S. Lashgari is supported  by NSF Grants CAREER 0953117, CCF-1161720, NETS-1161904, and ONR award N000141310094.

This work has been presented in part at the IEEE International Symposium on Information Theory 2014 ~\cite{OursISIT2014} and IEEE Globecom 2014 \cite{OursGlobecom}.
}
}

\begin{document}

\title{Blind MIMOME Wiretap Channel with Delayed CSIT}

\maketitle

\begin{abstract}
We study the Gaussian MIMOME wiretap channel 
 where a  transmitter  wishes to communicate a confidential message to a  legitimate receiver in the presence of  eavesdroppers, while the eavesdroppers  should not be able to decode the confidential message.
Each node in the network is equipped with arbitrary number of antennas.
Furthermore, 
channels are time varying, and there is no  channel state information available at the transmitter (CSIT) with respect to  eavesdroppers' channels; and transmitter only has access to delayed CSIT of the channel to the legitimate receiver.
The secure degrees of freedom (SDoF) for such network  has only been characterized for special cases, and is unknown in general.
We completely characterize the SDoF of this network for all antenna configurations.
In particular, we strictly improve the state-of-the-art achievable scheme for this network by proposing  more efficient artificial noise alignment at the receivers.
Furthermore, 
we develop a tight upper bound by 
utilizing 4 important inequalities that provide lower bounds on the received signal dimensions at receivers which supply delayed CSIT or no CSIT, or at a collection of receivers where some  supply no CSIT. These inequalities together allow for analysis of signal dimensions in networks with asymmetric CSIT; and as a result, we present a converse proof that leads to characterization of SDoF for all possible antenna configurations.
\end{abstract}

\section{Introduction}\label{introduction}
Wiretap channel is one of the canonical settings in the information-theoretic study of secrecy in wireless networks.
It consists of a transmitter that wishes to communicate a secret message to a legitimate receiver in the presence of   eavesdropper(s) that should not decode the confidential message. There has been a large amount of work on this problem, and its secrecy capacity has been determined in several configurations (e.g., \cite{Wyner, Csiszar, ITsecurity, GaussianWTp}). In particular, the secrecy capacity of the Gaussian wiretap channel is characterized in \cite{GaussianWTp}, and it is known that if the channel to the legitimate receiver is ``less noisy'' than the channel to the eavesdropper, then a positive rate of secret communication is achievable.


However, the secrecy capacity of the Gaussian wiretap channel does not scale with the available transmit power, i.e., the secure degrees of freedom (SDoF)  of Gaussian wiretap channel is zero.
 This has motivated 
the utilization of helping jammers and multi-antenna transmitters in  networks to increase the achievable SDoF (e.g.  \cite{Yener, Wornell, AnA, Ulukus, SDoFWTpBlind, Ravi, motahari, Zaidi1, Zaidi2, MIMOBCblind, Khisti}). In particular, it has been shown in \cite{Ulukus} that the SDoF of wiretap channel with a helping jammer (i.e. cooperative jamming) in a wireless setting in which  the channels remain constant is $\frac{1}{2}$.
This work has also been extended in \cite{SDoFWTpBlind} to the case that transmitters have no knowledge of channels to the eavesdropper (i.e., blind cooperative jamming), and it has been shown that 
even if transmitters have no eavesdropper channel state information at the transmitter (CSIT), the same SDoF can be achieved.
Furthermore, the SDoF of four fundamental one-hop wireless networks  has been studied in \cite{Ulukus}: Gaussian wiretap channel,
Gaussian broadcast channel with confidential messages, Gaussian interference channel with confidential messages, and Gaussian multiple access wiretap channel. They have provided  achievability results based on  Real Interference Alignment.
However, the above results rely on the assumption that  channels are constant, and do not change over time.

Secure communication over networks with time-varying channels (i.e. ergodic channels) has  been considered in some prior works in the literature \cite{poor, SDoFX, MIMOWTP, BlockFading,SDOFmISOBCalt}.
In particular, in \cite{poor}, achievability results for secure degrees of freedom of K-user interference channel with instantaneous channel state information at the transmitters (CSIT) were presented.
Moreover,  SDoF of wireless X networks has been studied in \cite{SDoFX}. 
Nevertheless, the results in these works heavily rely on the assumption that the transmitters have perfect and instantaneous access to the channel state information, which is not realistic in scenarios where channel coherence time is very small; because  by the time that channel state information is fed back  to the transmitter, the channel coefficients have already changed.
Therefore, there have been follow up works  that focus on studying SDoF for settings in which only delayed CSIT is available.
In particular, in \cite{MIMOWTP} Yang et al. have considered the Gaussian MIMO wiretap channel with delayed CSIT; and they have characterized the SDoF of such network for arbitrary number of antennas.
However, they assume that the eavesdropper supplies the transmitter with perfect delayed CSIT, which in most scenarios is not a realistic assumption. For the case where no eavesdropper CSIT is available, \cite{MIMOWTP} provides lower bounds on the SDoF which only match their respective upper bounds for specific network configurations, and the SDoF is in general unknown.

In this work, we focus on the time-varying wiretap channel in which  channels are changing over time; and we consider a multi-antenna transmitter with a multi-antenna legitimate receiver, and  arbitrary number of eavesdroppers, each equipped with multiple antennas (MIMOME wiretap channel). 
The transmitter
is \emph{blind} with respect to the state of channels to the eavesdroppers, and
only has access to \emph{delayed} channel state information (CSIT) of the legitimate receiver.
We consider the case where all nodes in the network are potentially equipped with multiple antennas, which is referred to as ``blind MIMOME wiretap channel with delayed CSIT".

For blind MIMOME wiretap channel with delayed CSIT,  SDoF has only been previously characterized for the special case where the number of transmit antennas is less than either the legitimate receiver or an eavesdropper \cite{MIMOWTP}.
For such case, a single-phase achievable scheme can achieve the SDoF, during which the transmitter sends a number of artificial noise symbols, equal to the number of antennas at the largest eavesdropper, and some information symbols in each time slot. 
However, SDoF is unsolved for other cases.
In particular, the  lower bounds on SDoF derived in the literature for the general case  do not match the existing upper bounds, which also hold for the setting in which eavesdroppers supply delayed CSIT \cite{MIMOWTP}.
In this paper, we completely characterize the SDoF for blind MIMOME wiretap channel with delayed CSIT for all antenna configurations.
In particular, we improve the state-of-the-art achievable schemes; and we provide tight upper bounds on the SDoF which capture the impact of having no eavesdropper CSIT on secure communications.

In our proposed achievable scheme the transmitter  transmits artificial noise symbols in order to perform two tasks simultaneously:  first, the artificial noise signals span the entire received signal space at the eavesdroppers to completely drown the confidential message in noise at the eavesdroppers.
Second,  artificial noise signals are aligned into a smaller linear subspace at the legitimate receiver in order to occupy less signal dimensions and leave some room for the confidential message to be decoded.\footnote{Artificial noise Alignment was introduced in \cite{AnA} to mask the confidential message in the artificial noise at the undesired receivers.}
Our achievable scheme performs these two tasks by utilizing the delayed CSIT provided by the legitimate receiver in a two-phase transmission scheme.
For settings in which the legitimate receiver has less antennas than an eavesdropper, our proposed achievable scheme allows for more efficient artificial noise alignment at the eavesdroppers by 
spending less  
artificial noise equations for retransmission, hence improving the achieved SDoF.



The converse proof is based on 4 main  lemmas.
Each lemma presents an inequality which provides a lower bound on the received signal dimension at a receiver which supplies a certain type of CSIT. 
These inequalities provide the essential tools for analyzing the received signal dimensions at different receivers in blind MIMOME wiretap channel with delayed CSIT.
In particular, Least Alignment Lemma, which generalizes the result in \cite{Davoodi} to the MIMO setting,  states that if two  receivers in a network have the same number of antennas and one of the receivers supplies  no CSIT,  the least amount of alignment will occur at that receiver, meaning that transmit signals will occupy the maximal signal dimensions at that receiver.

Moreover, Lemma \ref{ERI DCSIT} and Lemma \ref{ERI nCSIT} provide lower bounds on the received signal dimensions at  receivers which supply delayed and no CSIT, respectively.
Finally, Lemma \ref{Claimlemma} provides a lower bound on the received signal dimensions at a collection of  receivers, where some receivers  supply no CSIT.
The intuition behind the converse proof has been obtained by analyzing linear signal dimensions when transmitter is employing linear encoding schemes \cite{OursGlobecom}. 
The techniques used in the converse proof allow for a better understanding of the fundamental  trade-offs between the signal dimensions received at different receivers in networks with heterogeneous CSIT, where different receivers supply different types of CSIT.

{\bf Notation.} We use small letters (e.g. $ x$) for scalars, arrowed letters (e.g. $\vec x$) for vectors,  capital letters (e.g. $X$) for matrices, and  calligraphic font (e.g. $\mathcal X$) for sets.
For any scalar variable $x$  we use $[x]^+$ to denote $max(x,0)$.
Moreover, for a random vector $\vec x$, we denote its covariance matrix by $K_{\vec x}$.
Landau notation $x(n) = o(n)$ is used to denote $\lim_{n\to\infty} \frac{x(n)}{n}=0$.
We use $\det[A]$  to denote determinant of matrix $A$; and $A^H$ denotes Hermitian transpose of matrix $A$.
$I_{m}$ denotes the identity matrix of size $m\times m$; and $A \otimes B$ denotes the Kronecker product of matrices $A,B$.
$[x_1; x_2; \ldots ; x_n]$ denotes a $n\times 1$ vector with the i-th element being $x_i$.

\section{System Model and Main Results} \label{model}

\begin{figure}[h!]
\centering
\includegraphics[scale=.25, trim= 10mm 10mm 10mm 10mm]{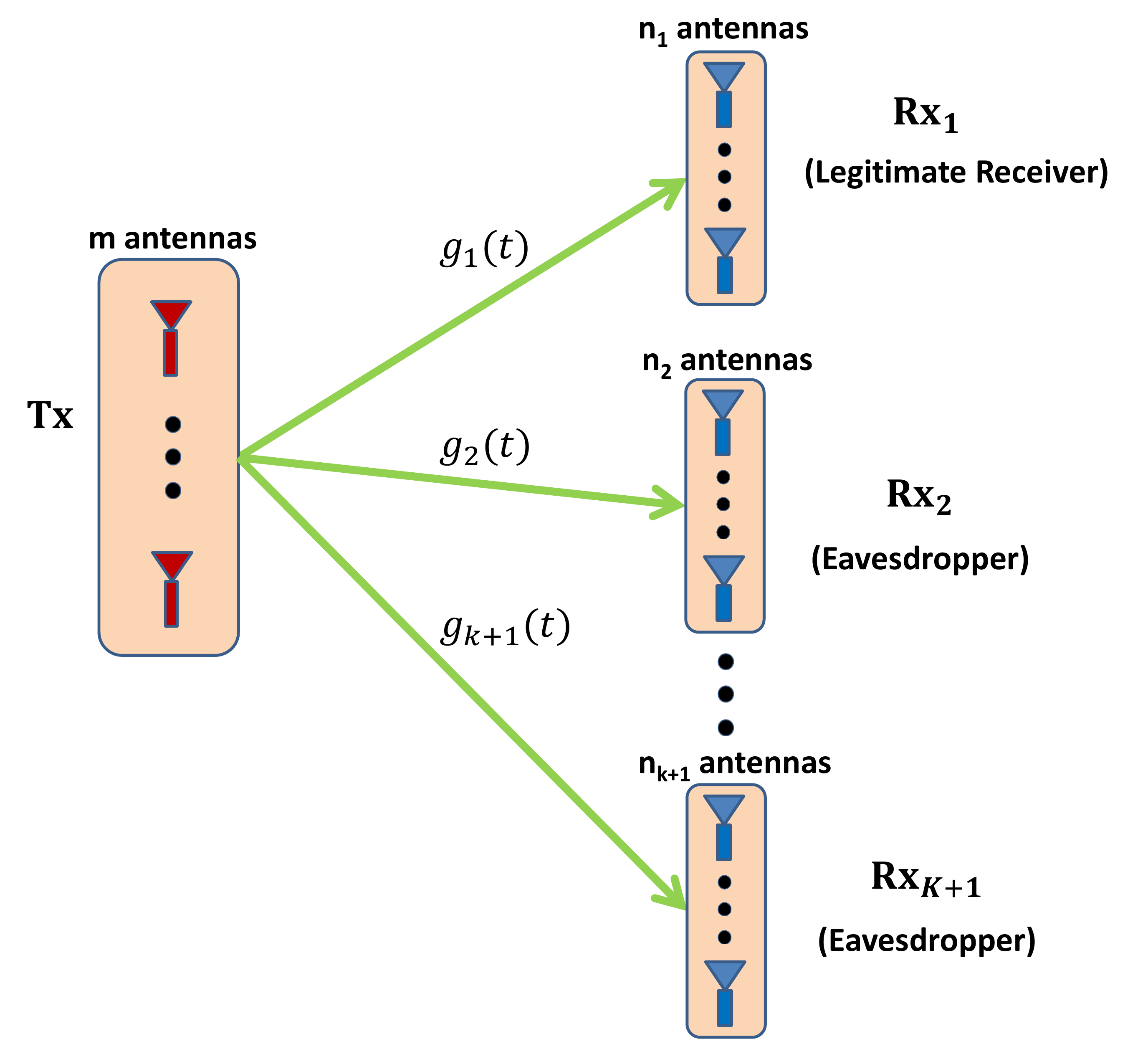}\\
\caption{
Network configuration for the blind MIMOME wiretap channel with $k$ eavesdroppers, where $\text{Rx}_2,\ldots, \text{Rx}_{k+1}$ do not supply any CSIT, and $\text{Rx}_1$ only supplies delayed CSIT.
}\label{wiretap}
\end{figure}

We consider the Gaussian multiple-input multiple-output multiple-eavesdropper (MIMOME) wiretap channel depicted in Figure \ref{wiretap}, which consists of
 a transmitter ($\Text{Tx}$) equipped with $m$ antennas and $k+1$ receivers $\text{Rx}_1,\text{Rx}_2, \\ \ldots , \Text{Rx}_{k+1}$, where $\text{Rx}_i$ ($i=1,\ldots ,k+1$) is equipped with  $n_i$ antennas ($m,n_1,\ldots , n_{k+1}\in \mathbb N$).
Throughout the paper we denote the maximum of $n_2, \ldots, n_{k+1}$ by $n_{max}$, and the corresponding receiver by $\text{Rx}_{max}$. 
In other words, $\text{Rx}_{max}$ is the eavesdropper with the most number of antennas ($n_{max}$ antennas).\footnote{If there are multiple eavesdroppers with $n_{max}$ antennas, we consider the eavesdropper with smallest index to be $\text{Rx}_{max}$.}
$\text{Tx}$ has a secret message for $\text{Rx}_1$ (legitimate receiver), while $\text{Rx}_2, \ldots , \Text{Rx}_{k+1}$ are eavesdroppers.

The received signal at $\text{Rx}_{j}$ ($j\in \{1,\ldots , k+1\}$) at time $t$ is given by
\begin{equation}
\vec { {y}}_{j}(t)={g}_{j}(t)\vec{{x}}(t)+\vec{{z}}_{j}(t),
\end{equation}
where
$\vec{{x}}(t) \in \mathbb C^{ m} $ is the transmit signal vector of $\text{Tx}$;
${g}_{j}(t) \in \mathbb C^{n_j\times m}$ indicates the channel matrix from 
$\text{Tx}$ to $\text{Rx}_{j}$;
and $\vec{{z}}_{j}(t) \sim \mathcal C \mathcal N(0,I_{n_j})$.
The channel coefficients  comprising the channel matrix ${g}_{j}(t)$ are i.i.d, and also i.i.d. across time,  antennas, and receivers, and they are drawn from a continuous distribution, where the absolute value of each element of ${g}_{j}(t)$ is bounded by a large number $d_{max}$.
We denote by ${ \mathcal {G}}(t) \triangleq  \{g_1(t),\ldots , g_{k+1}(t) \}$ the set of all  channel coefficients at time $t$.
In addition, we denote by ${\mathcal {G}}^n$ the set of all channel coefficients from time 1 to $n$, i.e., 
$${\mathcal {G}}^n \triangleq \{{g}_{j}(t): j \in \{1,\ldots ,k+1\}, t\in \{1,\ldots,n\}  \}.$$

Moreover, we use the following notation throughout the paper: 
 $$\vec{{x}}^t  \triangleq \begin{bmatrix}
    \vec{{x}}(1) \\ \vdots \\     \vec{{x}}(t)  \end{bmatrix} , \vec{{y}}_j^t  \triangleq \begin{bmatrix}
    \vec{{y}}_j(1) \\ \vdots \\     \vec{{y}}_j(t)  \end{bmatrix} , \vec{{z}}_{j}^t  \triangleq \begin{bmatrix}
    \vec{{z}}_j(1) \\ \vdots \\     \vec{{z}}_j(t)  \end{bmatrix}, {G}_{j}^t  \triangleq  diag ({G}_{j}(1), \ldots, {G}_{j}(t)),$$ 
where $diag ({G}_{j}(1), \ldots, {G}_{j}(t))$ is the block diagonal matrix which has ${G}_{j}(1), \ldots, {G}_{j}(t)$ on its diagonal.
Similarly,
 $$\vec{{x}}^{t_0:t_1} \triangleq \begin{bmatrix}
    \vec{{x}}(t_0) \\ \vdots \\     \vec{{x}}(t_1)  \end{bmatrix} , 
\vec{{y}}_j^{t_0:t_1}  \triangleq \begin{bmatrix}
    \vec{{y}}_j(t_0) \\ \vdots \\     \vec{{y}}_j(t_1)  \end{bmatrix} , 
\vec{{z}}_{j}^{t_0:t_1} \triangleq \begin{bmatrix}
    \vec{{z}}_j(t_0) \\ \vdots \\     \vec{{z}}_j(t_1)  \end{bmatrix}, 
{G}_{j}^{t_0:t_1} \triangleq  diag ({G}_{j}(t_0), \ldots, {G}_{j}(t_1)).$$ 

The transmitter obeys an average power constraint, $\frac{1}{n}E\{|| \vec { x}^n||^2 \}\leq p$.
We assume  delayed channel state information at the transmitter (CSIT) with respect to the channel to the legitimate receiver ($\Text{Rx}_1$); however, the transmitter does not have knowledge of the channels to the eavesdroppers.
In other words, at  time $t$, only  $ G_1^{t-1} $ is known precisely to the transmitter; and the transmitter only knows a probability distribution for values of the channels to the eavesdroppers, where we denote the maximum value of such distribution by $f_{max}$, where $f_{max}=o(\log p)$.

\begin{definition} \label{coding}
A code for a communication block length of $n$ for the blind MIMOME wiretap channel with delayed CSIT consists of:
\begin{itemize}
\item
A sequence of encoders, $f^{(n)} = (f_1^{(n)}, \ldots , f_n^{(n)})$, where at time $t$,
\begin{align*}
& {\vec x} (t) = f_{t}^{(n)}(  W, { G_1}^{t-1} ),  
\end{align*}
where message $W$ is uniformly distributed over $\{1,2,\ldots, |\mathcal W(n)|\}$.

\item
The corresponding decoder  $ F^{(n)}$ at $\text{Rx}_1$, where
\begin{align*}
&\hat W =  F^{(n)} ( { { \vec y}}_{1}^n, { \mathcal {G}}^{n}  ).
\end{align*}

\item
The error probability of communication is defined to be
\begin{align*}
& p_e^{(n)}= \Pr(W\ne \hat W).
\end{align*}

\end{itemize}

\end{definition}

Based on the above definition, we now define the  secure degrees of freedom (SDoF) of the blind MIMOME wiretap channel with delayed CSIT.
\begin{definition}
$d$ secure degrees of freedom are  achievable if there exists a sequence
of encoders and decoders  $\{ f^{(n)}, F^{(n)} \}_{n=1}^{\infty}$,
such that 
\begin{equation} \label{condition1}
\limsup_{n\to\infty} p_e^{(n)}=0,
\end{equation}
and 
\begin{equation} \label{condition2}
\liminf_{p\to\infty}  \liminf_{n\to\infty} \frac{\log |\mathcal W(n)|}{n\log p} \geq  d,
\end{equation}

and Equivocation condition is satisfied:
\begin{equation}
\limsup_{p\to\infty}  \limsup_{n\to\infty} \frac{  I(W ;  { { \vec y}}_{j}^n )   }{n\log p  } \stackrel{}{=} 0, \quad 2\leq j\leq k+1.\label{condition3}
\end{equation}

We define $\mathcal D$ to be the convex closure of the set of all achievable $d$'s.
We also define secure degrees of freedom ($\Text{SDoF}$) to be the supremum of all $d\in \mathcal D$.
\end{definition}

\begin{remark}
Equivocation condition in (\ref{condition3})  means that the prelog factor of the equivocation rate to eavesdroppers should vanish as $n\to\infty$.
The Equivocation condition (\ref{condition3}) is weaker than the condition $\limsup_{n\to\infty} \frac{I(W;\vec {\bf y}_{j}^n)}{n} \stackrel{}{=} 0, \quad 2\leq j\leq k+1$, considered in some prior works including \cite{Wolf, Narayan}. 
However, one can combine  our achievable schemes presented in Section \ref{achievabilityproof} for blind MIMOME  wiretap channel with delayed CSIT  with random binning \cite{Wyner} to satisfy the latter condition as well.
\end{remark}

For the problem of blind  MIMOME wiretap channel with delayed CSIT, Yang et  al.  ~\cite{MIMOWTP} 
have characterized  SDoF for the special case where one of the receivers has more antennas than the transmitter.
However, SDoF is in general unknown, and \cite{MIMOWTP} only provides lower bounds on SDoF for the general case.

The main result of this paper is the following theorem, which completely characterizes  SDoF for blind MIMOME wiretap channel with delayed CSIT for all antenna configurations by
 improving the best known achievable schemes and  providing tight upper bounds.

\begin{theorem}\label{maintheorem}
For the blind MIMOME wiretap channel with delayed CSIT, 
let $\bar m= min(m,n_1+n_{max})$ and $\bar n = min(n_1, n_{max})$.
Then, $\Text{SDoF} $ is characterized as follows:
\large
\begin{equation}
\Text{SDoF} = \Bigg \{
  \begin{tabular}{ccc}
  $[m-n_{max}]^{+}$ & if $m \leq max(n_1,n_{max})$ \\  
 $\frac{n_1(\bar m - n_{max})}{\bar m - n_{max} + \bar n}$  & if $m > max(n_1,n_{max}).$
  \end{tabular}
\end{equation}
\normalsize
\end{theorem}

Note that the SDoF of blind MIMOME wiretap channel with delayed CSIT does not decrease with increasing the number of eavesdroppers; rather, SDoF is only a function of $n_{max}$, the maximum number of antennas on a single eavesdropper.
As a special case, Theorem \ref{maintheorem} implies that the achievable scheme presented in \cite{Shamai} for blind MISO wiretap channel with delayed CSIT, which achieves $\frac{1}{2}$, is indeed optimal.\footnote{This special case has also been independently studied in \cite{SDOFmISOBCalt}.}

In order to better understand how  the result compares with  prior works from the achievability perspective, Table \ref{comparetable} presents two classes of antenna configurations for which our results strictly improve the existing achievable schemes. 
In particular, the table provides an example for each case, and states the achieved secure degrees of freedom by both \cite{MIMOWTP} and  Theorem \ref{maintheorem}.
Thus, Theorem \ref{maintheorem} strictly improves the existing achievable schemes in  cases such as $n_1\leq n_{max} < m \leq n_1+ n_{max}$,  and $m \geq n_1+ n_{max}$, where $n_{max}>n_1$.
Moreover,  as we will see in  Section \ref{achievabilityproof},
we provide a  single unified achievable scheme  for all antenna configurations which satisfy  $max(n_1,n_{max}) < m $.

\renewcommand{\arraystretch}{2}
\begin{table*}[t]
\centering
\begin{tabular}{|c|c|c|c|}
\hline
$\text{Antenna Configuration}$ &Example & Achieved SDoF in \cite{MIMOWTP} &SDoF \\
\hline

 $n_1\leq n_{max} < m \leq n_1+n_{max}$  & 
$m=4, n_1=2, n_{max}=3$, illustrated in Fig. \ref{MIMOCase3} & $\frac{n_1(m-n_{max})}{m}$ & $\frac{n_1(m-n_{max})}{m-n_{max} + n_1}$ \\
\hline 
$n_1\leq n_{max}, m>n_1 + n_{max}$  & $m=3, n_1=1, n_{max}=2$, illustrated in Fig. \ref{MIMOCase4} & $\frac{n_1^2}{n_1+n_{max}}$   & $\frac{n_1}{2}$\\[1mm]
\hline
\end{tabular}
\vspace{3mm}
\caption{  Comparison of achievability results for 2 different classes of antenna configuration
}
\label{comparetable}
\end{table*}

Theorem \ref{maintheorem} captures the fundamental impact of availability of delayed CSIT from the legitimate receiver and no eavesdropper CSIT. 
More specifically, 
by comparing SDoF of \emph{blind} MIMOME wiretap channel with delayed CSIT with that of 
MIMOME wiretap channel with delayed CSIT \cite{MIMOWTP},
as done in Table \ref{compareSDoFsTable},
 one can note that
 SDoF strictly decreases when there is no eavesdropper CSIT available and $max(n_1,n_{max}) < m$. 

\renewcommand{\arraystretch}{2}
\begin{table*}[t]
\centering
\begin{tabular}{|c|c|c|c|}
\hline
$\text{Network}$ &
\begin{tabular}{ccc}
 SDoF when\\
$ m \leq max(n_1,n_{max})$
  \end{tabular}
&
\begin{tabular}{ccc}
 SDoF when\\
$max(n_1,n_{max}) < m \leq n_1+n_{max}$
  \end{tabular}
&
\begin{tabular}{ccc}
 SDoF when\\
$m>n_1 + n_{max}$
  \end{tabular}
\\
\hline

\begin{tabular}{ccc}
 MIMOME WTP\\
with delayed CSIT
  \end{tabular}  & 
$[m-n_{max}]^{+}$ & $\frac{n_1m(m - n_{max})}{n_1n_{max} + m(m - n_{max})}$ & $\frac{n_1(n_1 + n_{max})}{n_1 + 2n_{max}}$ \\
\hline 
\begin{tabular}{ccc}
Blind  MIMOME WTP\\
with delayed CSIT
  \end{tabular}  &
$[m-n_{max}]^{+}$ & $\frac{n_1(m - n_{max})}{m - n_{max}+min(n_1,n_{max}) }$ & $\frac{n_1^2}{n_1 + min(n_1,n_{max})}$\\
\hline
\end{tabular}
\vspace{3mm}
\caption{  Comparison of SDoF for two networks: MIMOME wiretap channel with delayed CSIT, and blind MIMOME wiretap channel with delayed CSIT
}
\label{compareSDoFsTable}
\end{table*}

In the following sections  we provide the proof for  Theorem \ref{maintheorem}, and explain the key ideas behind the proof.

\section{proof of Achievability} \label{achievabilityproof}

In this section we present the achievable schemes for all  antenna configurations.
At a high level, our scheme  transmits two types of symbols: information symbols, which together constitute the confidential message, and artificial noise symbols. 
By using an appropriate linear precoder and utilizing delayed CSIT supplied by the legitimate receiver, we align artificial noise symbols into a smaller linear subspace at the legitimate receiver so that some room is left for decoding information symbols, while allowing artificial noise to occupy the whole received signal space at the eavesdroppers. 
This completely drowns information symbols in artificial noise at the eavesdroppers so that eavesdroppers will not be able to decode them, while it allows the legitimate receiver to successfully decode the information symbols.

Throughout the presentation of the achievable schemes, for simplicity and without loss of generality we assume that 
$n = n'b$, where $b$ is the block length of communication for our scheme, and $n'$ is the number of blocks. 
In fact, we repeat the same transmission scheme for all blocks.
We first present the achievable scheme for the case of $m \leq max(n_1,n_{max})$.\footnote{The achievable scheme for the case  $m \leq max(n_1,n_{max})$ is presented in \cite{MIMOWTP};  however,  we  state it here for completion and because its analysis serves as an introduction to analysis of  the case   $m > max(n_1,n_{max})$.}

\subsection{Case of $m \leq max(n_1,n_{max})$}
Note that for the case where $m\leq n_{max}$, Theorem \ref{maintheorem} suggests that  $\Text{SDoF} = 0$; so 
there is nothing to prove on the achievability side.
Thus, let us consider the case where $n_{max} < m\leq n_1$.
In this case we will show that $d= m-n_{max}$ secure degrees of freedom is achievable. 
In other words,  we show how to securely deliver $m- n_{max}$ information symbols to $\Text{Rx}_1$ in each time slot ($b=1$).
In particular, in every time slot,
each of the first $n_{max}$ transmit antennas sends a distinct artificial noise symbol, while each of the antennas with index $n_{max}+1, \ldots, m-1, m$ sends a distinct new information symbol.
Consequently, $\Text{Rx}_1$ recovers all symbols almost surely, including the $m- n_{max}$ information symbols, since it receives $n_1$ equations  in $m$ unknowns where $m\leq n_1 $, hence satisfying (\ref{condition2}). 
By using an appropriate code for the  Gaussian MIMO channel between $\text{Tx}$ and $ \text{Rx}_1$,
when the block length of communication  goes to infinity (i.e. $n\to\infty$), the error of communication goes to zero, satisfying (\ref{condition1}).

Moreover, 
the eavesdroppers cannot decode any of the information symbols  because
$\text{Rx}_{max}$ essentially receives $n_{max}$ equations regarding $n_{max}$ artificial noise symbols (undesired symbols) and $(m-n_{max})$ information symbols.
Hence, the information symbols are completely drowned in artificial noise and Equivocation condition (\ref{condition3}) is satisfied.

After providing the intuitive reason why the achievable scheme satisfies  conditions (\ref{condition2})-(\ref{condition3}), we now rigorously prove that under the proposed scheme, conditions (\ref{condition2})-(\ref{condition3}) are satisfied.
To this aim, we  state a lemma that will be useful in the proof of achievability. 
\begin{lemma} \label{entropyrank}
For a fixed matrix A,
\begin{equation}
\lim_{p\to \infty} \frac{\log \det[I+ pAA^{H}]}{\log p} = \text{rank}[A].
\end{equation}
\end{lemma}
Proof of Lemma \ref{entropyrank} follows from straightforward linear algebra and can be found in \cite{MIMOWTP}.

We first specify the transmit signals, and then use them to show that conditions (\ref{condition2})-(\ref{condition3}) are satisfied.
At time slot $t$, 
 $\vec u_{n_{max}\times 1}  \in \mathcal C \mathcal N(0,\frac{p}{ m} I_{ n_{max}})$ denotes the vector of  artificial noise  symbols, which are transmitted by antennas $1, \ldots, n_{max}$, and the vector of information symbols by $\vec v_{(m- n_{max})\times 1} \in \mathcal C \mathcal N(0,\frac{p}{m} I_{(m- n_{max})})$, which are transmitted by antennas $n_{max}+1, \ldots, m$.
As a result,
\begin{align}
\vec x =  \begin{bmatrix}
    \vec u_{n_{max}\times 1}   \\  \vec v_{(m- n_{max})\times 1}\end{bmatrix} , 
\qquad  K_{\vec x} = \frac{p}{m} I_{m}, 
\qquad K_{\vec x | \vec v} =  \frac{p}{m}
 \begin{bmatrix}
    I_{n_{max}} & 0  \\  0& 0\end{bmatrix}  . 
\end{align}
Therefore, for any $j\in\{1,2,\ldots , k+1\}$,
\begin{align}
\lim_{p\to\infty} \frac{I(\vec v ; \vec y_j | \mathcal G)}{\log p}  &= \lim_{p\to\infty} E_{\mathcal G} \{ \frac{h(\vec y_j | \mathcal G ) - h(\vec y_j | \vec v, \mathcal G )}{\log p}  \}   = \lim_{p\to\infty} E_{\mathcal G} \{ \frac{ \log\det [I+  G_j K_{\vec x}G_j^H] -  \log\det [I+  G_j K_{\vec x | \vec v}G_j^H]  }{\log p}  \}  \nonumber\\
& \stackrel{(a)}{=} E_{\mathcal G} \{   \lim_{p\to\infty} \frac{ \log\det [I+  G_j K_{\vec x}G_j^H] -  \log\det [I+  G_j K_{\vec x | \vec v}G_j^H]  }{\log p}  \}  \nonumber\\
&\stackrel{}{=}    E_{\mathcal G} \{ \lim_{p\to\infty} \frac{ \log\det [I+ \frac{p}{m} G_j G_j^H] }{\log p} -  \lim_{p\to\infty} \frac{ \log\det [I+ \frac{p}{m} G_j  \begin{bmatrix} I_{n_{max}} & 0 \\  0 &   0_{(m-n_{max})} \end{bmatrix}   G_j^H] }{\log p}  \}  \nonumber\\
&\stackrel{(\text{Lemma } \ref{entropyrank})}{=}    E_{\mathcal G} \{ \text{rank}[G_j] - \text{rank}[G_j \begin{bmatrix} I_{n_{max}} & 0 \\  0 &   0_{(m-n_{max})} \end{bmatrix}]    \}  ,  \label{1stcaseanalysis}
\end{align}
where (a) is due to Dominated Convergence Theorem.
Furthermore, note that for a random channel realization $\mathcal G$,
\begin{align}
\text{rank}[G_j] \stackrel{a.s.}{=} min(n_j,m), \qquad   
\text{rank}[G_j \begin{bmatrix} I_{n_{max}} & 0 \\  0 &   0_{(m-n_{max})} \end{bmatrix}]  \stackrel{a.s.}{=} min(n_j,n_{max}).\label{1stcaseanalysis2}
\end{align}
Hence, since we are considering the case where $n_{max} < m\leq n_1$,
 by (\ref{1stcaseanalysis}), (\ref{1stcaseanalysis2}), we obtain
\begin{align}
&\lim_{p\to\infty} \frac{I(\vec v ; \vec y_1 | \mathcal G)}{\log p} = m-n_{max},\\
&\lim_{p\to\infty} \frac{I(\vec v ; \vec y_j | \mathcal G)}{\log p} = 0, \qquad j=2,\ldots,k+1,
\end{align}
which prove that conditions (\ref{condition2}),(\ref{condition3}) are satisfied. 
Hence, the proposed achievable scheme achieves   $[m-n_{max}]^{+}$ secure degrees of freedom when $m \leq max(n_1,n_{max})$.

\begin{figure*}
\centering
\includegraphics[scale=.4, trim= 10mm 10mm 10mm 10mm]{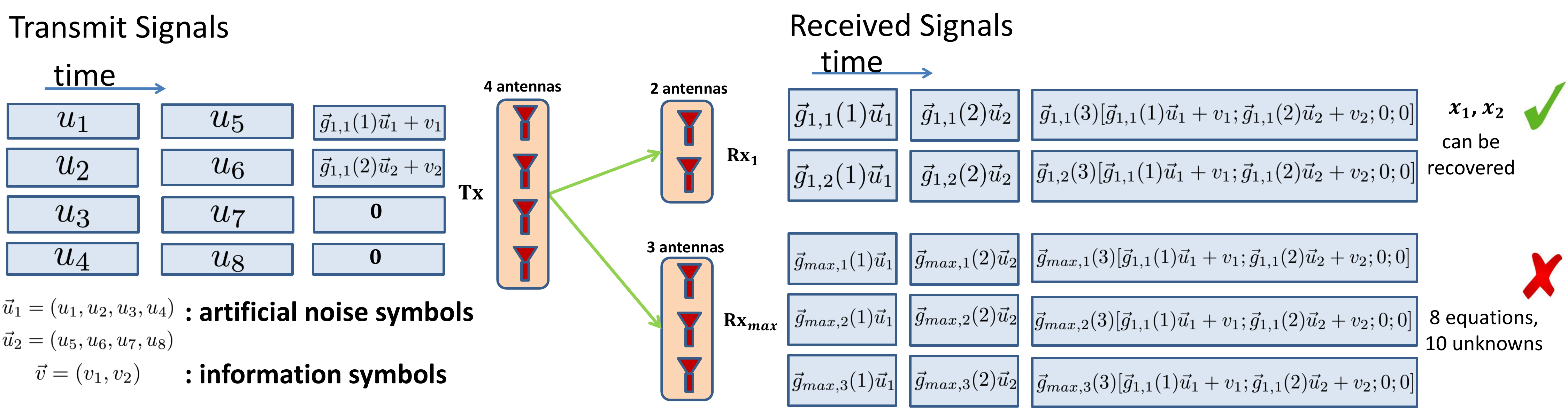}\\
\caption{
The achievable scheme for  a simple network configuration that belongs to case where $n_1  \leq n_{max} , m\geq  n_1 +  n_{max}$.
The scheme delivers 2 symbol securely over 3 time slots, achieving SDoF of $\frac{2}{3}$. 
}\label{MIMOCase3}
\end{figure*}

\subsection{Case of  $m > max(n_1,n_{max})$}
In this case the scheme will securely deliver $n_1 (\bar m- n_{max})$ information symbols over $\bar m -n_{max} + \bar n$ time slots ($b= \bar m -n_{max} + \bar n$).
The scheme is presented in two phases. The first phase takes $\bar n$ time slots, during which only  artificial noise symbols are transmitted. Then, during the second phase, which takes $\bar m -n_{max} $ time slots, some of  artificial noise equations are retransmitted together with information symbols in such a way that they completely mask the information symbols at the eavesdroppers, while the information symbols can be recovered at the legitimate receiver.
Since throughout the description of our achievable scheme and its analysis only $\bar m$ transmit antennas are used at any point in time, we only focus on the first $\bar m$ transmit antennas for the sake of simplicity and ignore the rest.
Further, we implicitly  consider the proper scaling which is needed to satisfy the power constraint.
We now present the details of our scheme.

\underline{Phase 1:} 
This phase takes $\bar n$ time slots.
At $t=1,2,\ldots , \bar n$, each of the $\bar m$ transmit antennas  sends a  new artificial noise symbol. 
Thus,  since channel coefficients are i.i.d. and drawn from a continuous distribution,
what $\Text{Rx}_1$ receives on its first $(\bar m-n_{max})$ antennas are almost surely linearly independent of equations received by antennas of $\Text{Rx}_{max}$, hence not recoverable by $\Text{Rx}_{max}$.  
Similar result holds for other eavesdroppers.
Hence, by the end of phase 1, $\Text{Rx}_1$ obtains $\bar n (\bar m-n_{max})$ linearly independent noise equations that are not recoverable  by eavesdroppers.

\underline{Phase 2:} 
This phase takes $\bar m -n_{max}$ time slots.
In each time slot $t\in \{\bar n+1, \ldots,\bar m- n_{max} + \bar n \}$,  transmit signals by the $\bar m$ transmit antennas are as follows:
\begin{equation}
\vec x(t) =  
\left [  \begin{tabular}{ccc}
  $\vec u'_{\bar n}(t)$ \\ $\vec 0_{\bar m - \bar n}$  \end{tabular} \right ] 
+ \left [  \begin{tabular}{ccc}
  $\vec v_{n_1}(t)$ \\ $\vec 0_{\bar m - n_1}$  \end{tabular} \right ],
\end{equation}
where $\vec u'_{\bar n}(t) $ is a vector  comprised of $\bar n$ linearly independent artificial noise equations known  by $\Text{Rx}_1$ (ignoring AWGN), which are not recoverable by eavesdroppers, and are produced as the result of Phase 1 of the scheme. 
Note that these artificial noise equations are known to the transmitter by the end of Phase 1 because it has access to delayed CSIT of the legitimate receiver as well as all artificial noise symbols.
We will refer to  these artificial noise equations as artificial noise symbols for simplicity.
Moreover, $\vec v_{n_1}(t) $
 is the vector of   information symbols.

As a result of this transmission scheme, in each time slot of Phase 2, $\Text{Rx}_1$ cancels the transmitted artificial noise symbols from its received signal, and recovers the $n_1$ desirable information symbols. Therefore, $\Text{Rx}_1$ recovers $n_1( \bar m -n_{max})$ information symbols in total, satisfying (\ref{condition2}).

In order to  explain why  Equivocation condition (\ref{condition3}) is satisfied , we focus on $\text{Rx}_{max}$ as the argument  is similar for other eavesdroppers.
We consider the two antenna configurations $\bar n = n_1 \leq n_{max}$, and $\bar n = n_{max} \leq n_1$.
In the case of $\bar n = n_1 \leq n_{max}$, each active antenna is sending one information symbol plus an artificial noise symbol which is not known to $\text{Rx}_{max}$. 
Therefore,  $\text{Rx}_{max}$ cannot  recover any of the information symbols.
Moreover, in the case of $\bar n = n_{max} \leq n_1$,  the transmitter is sending $n_{max}$ artificial noise symbols in each time slot which are not known to $\text{Rx}_{max}$. 
Thus, since $\text{Rx}_{max}$ has $n_{max}$ antennas, the transmitted artificial noise symbols span the entire space of received signals at $\text{Rx}_{max}$ and do not allow for any information symbol to be recovered, satisfying (\ref{condition3}).

\begin{figure*}
\centering
\includegraphics[scale=.4, trim= 10mm 10mm 10mm 10mm]{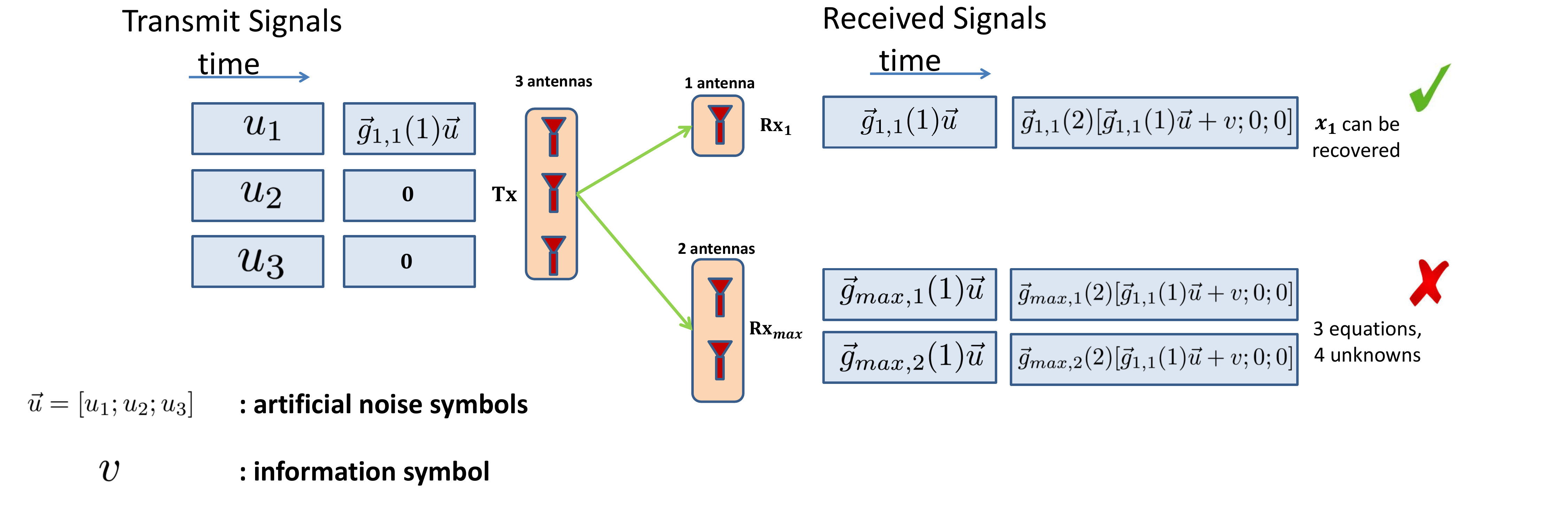}\\
\caption{
The achievable scheme for  a simple network configuration that belongs to case where $n_1  \leq n_{max} , m\geq  n_1 +  n_{max}$.
The scheme delivers 1 symbol securely over 2 timeslots, achieving SDoF of $\frac{1}{2}$. 
}\label{MIMOCase4}
\end{figure*}

Hence, overall $n_1(\bar m-n_{max})$ information symbols are delivered securely to $\Text{Rx}_1$ over $\bar m - n_{max} + \bar n$ time slots, and the scheme achieves $\frac{n_1(\bar m - n_{max})}{\bar m - n_{max} + \bar n}$ secure degrees of freedom. 
Thus, the proposed achievable scheme strictly improves the existing achievable schemes  \cite{MIMOWTP} for the case $n_1<n_{max} < m \leq n_1+ n_{max}$, illustrated in Figure \ref{MIMOCase3},
and $m \geq n_1+ n_{max}$, where $n_{max}>n_1$, as illustrated in Figure \ref{MIMOCase4}.

After explaining why the proposed achievable scheme achieves $\frac{n_1(\bar m - n_{max})}{\bar m - n_{max} + \bar n}$ secure degrees of freedom, we now prove that the conditions (\ref{condition2}),(\ref{condition3}) are indeed satisfied.
Let 
$ \vec u \sim \mathcal C \mathcal N (0, \frac{p}{\bar m}I_{\bar n\bar m} ),
\vec v \sim \mathcal C \mathcal N (0, \frac{p}{\bar m}I_{ n_1( \bar m - n_{max})}    ).$
The transmit signal is as follows:
\small
\begin{align}
 \vec x^{\bar m -n_{max} + \bar n} = \begin{bmatrix} I_{\bar n\bar m} & 0_{\bar n \bar m \times n_1(\bar m-n_{max})} \\  A &    
I_{\bar m-n_{max}}  \otimes \left [  
  \begin{tabular}{ccc}
 $ I_{n_1}$ & 0  \\  0 & $0_{\bar m -n_1} $   \end{tabular}
\right ] 
\end{bmatrix}
\begin{bmatrix}
\vec u \\
\vec v
\end{bmatrix},
\end{align}
where
\begin{align}
A \triangleq \begin{bmatrix} 
A(\bar n+1)\\
\vdots \\
A(\bar m - n_{max}+ \bar n)
\end{bmatrix},
\quad \text{ and } \quad 
A(t) \triangleq \begin{bmatrix} 
\vec g_{1,t-\bar n}(1) & 0 & \ldots & 0 \\
0 & \vec g_{1,t-\bar n}(2)  & \ldots & 0 \\
\vdots && \ddots &\vdots \\
 0 & \ldots & 0 & \vec g_{1,t-\bar n}(\bar n)  \\
&0_{(\bar m-\bar n)\times \bar n \bar m }&&
\end{bmatrix},
\end{align}
\normalsize
where $\vec g_{1,i}(t)$ is the channel vector comprised of coefficients of channels between transmitter and the $i$-th receive antenna of $\text{Rx}_1$ at time $t$ . 
Hence, the received signal at $\text{Rx}_j$ ($j=1,\ldots , k+1$) is 
\begin{align}
 \vec y_j^{\bar m -n_{max} + \bar n} = \begin{bmatrix} G_j^{\bar n} && 0_{\bar n n_j \times n_1(\bar m-n_{max})} \\  G_j^{\bar n+1: \bar m -n_{max}+\bar n} A &&   
G_j^{\bar n+1: \bar m -n_{max}+\bar n}   I_{\bar m-n_{max}}  \otimes \left [  
  \begin{tabular}{ccc}
 $ I_{n_1}$ & 0  \\  0 & $0_{\bar m -n_1} $   \end{tabular}
\right ] 
\end{bmatrix}
\begin{bmatrix}
\vec u \\
\vec v
\end{bmatrix} + \vec z_j^{\bar m -n_{max} + \bar n}.
\end{align}

We now show that (\ref{condition2})-(\ref{condition3}) are satisfied.
Using similar steps as in (\ref{1stcaseanalysis}), we obtain
\begin{align}
\lim_{p\to\infty} \frac{I(\vec v; \vec y_j^{\bar m - n_{max} + \bar n })}{\log p} &= E_{\mathcal G^{\bar m - n_{max} + \bar n }}
\{
\text{rank} \begin{bmatrix} G_j^{\bar n} && 0_{\bar n n_j \times n_1(\bar m-n_{max})} \\  G_j^{\bar n+1: \bar m -n_{max}+\bar n} A &&   
G_j^{\bar n+1: \bar m -n_{max}+\bar n}   I_{\bar m-n_{max}}  \otimes \left [  
  \begin{tabular}{ccc}
 $ I_{n_1}$ & 0  \\  0 & $0_{\bar m -n_1} $   \end{tabular}
\right ] 
\end{bmatrix}\}  \nonumber\\
&\qquad  - E_{\mathcal G^{\bar m - n_{max} + \bar n }}
\{  \text{rank}
\begin{bmatrix} G_j^{\bar n} \\  G_j^{\bar n+1: \bar m -n_{max}+\bar n} A 
\end{bmatrix}
\}. \label{achproverank}
\end{align}
Due to the structure of $A$, 
\begin{align}
\text{rank}
\begin{bmatrix} G_j^{\bar n} \\  G_j^{\bar n+1: \bar m -n_{max}+\bar n} A   
\end{bmatrix}
\stackrel{a.s.}{=} \Bigg\{
  \begin{tabular}{ccc}
  $\text{rank} [G_j^{\bar n}]$, &$\qquad \text{for } j=1$   \\
 $\text{rank} [G_j^{\bar n}] + \text{rank} [G_j^{\bar n+1: \bar m -n_{max}+\bar n} A ]$, &$\qquad \text{for } j>1$
  \end{tabular} \label{numnoiseeqs}
\end{align}
and
\begin{align}
&\text{rank} \begin{bmatrix} 
G_j^{\bar n} && 0_{\bar n n_j \times n_1(\bar m-n_{max})} \\  
G_j^{\bar n+1: \bar m -n_{max}+\bar n} A &&   G_j^{\bar n+1: \bar m -n_{max}+\bar n}   I_{\bar m-n_{max}}  \otimes \left [  
  \begin{tabular}{ccc}
 $ I_{n_1}$ & 0  \\  0 & $0_{\bar m -n_1} $   \end{tabular}
\right ] 
\end{bmatrix}  \nonumber\\
& \stackrel{a.s.}{=} 
\Bigg\{
  \begin{tabular}{ccc}
  $\text{rank} [G_j^{\bar n}]+ \text{rank}\left[G_j^{\bar n+1: \bar m -n_{max}+\bar n}   I_{\bar m-n_{max}}  \otimes \left [  
  \begin{tabular}{ccc}
 $ I_{n_1}$ & 0  \\  0 & $0_{\bar m -n_1} $   \end{tabular} \right ] \right]$, &$\qquad \text{for } j=1$   \\
 $\text{rank} [G_j^{\bar n}] + \text{rank} [G_j^{\bar n+1: \bar m -n_{max}+\bar n} A ]$, &$\qquad \text{for } j>1$
  \end{tabular}
 \nonumber\\
& \stackrel{a.s.}{=} 
\Bigg\{
  \begin{tabular}{ccc}
  $\text{rank} [G_j^{\bar n}]+n_1(\bar m-n_{max})$, &$\qquad \text{for } j=1$   \\
 $\text{rank} [G_j^{\bar n}] + \text{rank} [G_j^{\bar n+1: \bar m -n_{max}+\bar n} A ]$, &$\qquad \text{for } j>1.$
  \end{tabular} \label{numtotaleqs}
\end{align}

Using  (\ref{achproverank})-(\ref{numtotaleqs}), one can readily see that conditions (\ref{condition2})-(\ref{condition3}) are satisfied.
Hence, the proposed achievable scheme achieves $\frac{n_1(\bar m - n_{max})}{\bar m - n_{max} + \bar n}$  secure degrees of freedom when $m > max(n_1,n_{max})$; and thus,
 achievability proof  is complete.

\section{Proof of Converse for Theorem \ref{maintheorem}}
In this section we present the converse proof for Theorem \ref{maintheorem}. 
Note that for any antenna configuration ($m,n_1,n_2,\\ \ldots, n_{k+1}$), if some of the eavesdroppers are removed from the network, $\Text{SDoF}$ will not decrease; and this is due to removing some of the Equivocation constraints on maximizing the secure rate.
Hence, to prove the converse we first remove all the eavesdroppers except $\Text{Rx}_{max}$ from the network.
We start by proving the converse for the case where $m\leq max(n_1,n_{max})$.\footnote{Proof of converse for the case when $m\leq max(n_1,n_{max})$ has been presented in \cite{MIMOWTP}; nevertheless, we provide the proof here for completion.}

\subsection{Proof of Converse for  $m\leq max(n_1,n_{max})$}
We first state a lemma that will be used throughout this section.

\begin{lemma} \label{ERI DCSIT}
Consider two receivers $\text{Rx}_1,\text{Rx}_{2}$ with $n_1,n_{2}$ antennas, where $\text{Rx}_2$ supplies delayed CSIT or no CSIT. 
For any fixed $n$, and any encoding strategy $ f^{(n)} $, 
 $$  \frac{h(\vec y_{2}^n | \mathcal G^n )}{min(m,n_{2})} + n\times o(\log p) \geq  \frac{ h(\vec y_{1}^n, \vec y_{2}^n | \mathcal G^n )}{min (  m, n_1 + n_{2}  )} .$$
\end{lemma}
Lemma \ref{ERI DCSIT} provides a lower bound on the received signal dimension at a receiver which does not supply perfect CSIT.
In other words, it relates the received signal dimension of a receiver which supplies delayed or no CSIT to that of another receiver.
Proof of  Lemma \ref{ERI DCSIT} follows from channel symmetry; and it can be found  in \cite{varanasi, MIMOWTP}.

We now prove the converse for $m\leq max(n_1,n_{max})$.
Suppose $d$ secure DoF is achievable.
Therefore, by Fano's inequality,
\begin{align}
n(R-\epsilon_n) \stackrel{}{\leq} & I(W;\vec y_1^n  | \mathcal G^n) \leq I(W;\vec y_1^n ,\vec y_{max}^n  | \mathcal G^n)     \nonumber\\
&\stackrel{(a)}{\leq}   h(\vec y_1^n ,\vec y_{max}^n  | \mathcal G^n)- h(\vec y_{max}^n  | W,\mathcal G^n) \nonumber \\
 &=    h(\vec y_1^n ,\vec y_{max}^n  | \mathcal G^n)- h(\vec y_{max}^n  | \mathcal G^n) + I(W ; \vec y_{max}^n  | \mathcal G^n)  \nonumber \\ 
 &\stackrel{(\ref{condition3})}{\leq}  h(\vec y_1^n ,\vec y_{max}^n  | \mathcal G^n)- h(\vec y_{max}^n  | \mathcal G^n) + n\times o(\log p)  \nonumber \\
& \substack{\text{(Lemma \ref{ERI DCSIT})} \\ \leq}  \frac{  m -min(m, n_{max})}{min(m, n_{max})}  \times h(\vec y_{max}^n  | \mathcal G^n) + n\times o(\log p) \nonumber\\
& =  \frac{[m- n_{max}]^+}{n_{max}}  \times h(\vec y_{max}^n  | \mathcal G^n) + n\times o(\log p) \nonumber\\
& \stackrel{}{\leq} \frac{[m- n_{max}]^+}{n_{max}}  \times nn_{max}\log p  + n\times o(\log p) \nonumber\\
& \stackrel{}{=} [m-n_{max}]^{+} \times n\log p  + n\times o(\log p), \nonumber
\end{align}
where (a) holds since $  h(\vec y_1^n |  \vec y_{max}^n , W, \mathcal G^n) \geq h(\vec y_1^n | \vec x^n,  \vec y_{max}^n ,W, \mathcal G^n) = h(\vec z_1^n)>0$.
Hence, by dividing both sides of the above inequality by $n\log p$ and taking the limit $n\to \infty$ and $p\to \infty$, the result follows.

\subsection{Proof of Converse for  $m> max(n_1,n_{max})$}
Before presenting the converse proof for $m> max(n_1,n_{max})$, we consider a notation that will be used throughout the proof.
Let $\text{Rx}_{max,1}$ denote the first $\bar n$ antennas on $\text{Rx}_{max}$ with the corresponding received signal of $\vec y_{max,1}^n$ over $n$ time slots.
Further, let $\text{Rx}_{max,2}$ denote the remaining $n_{max} - \bar n$ antennas on $\text{Rx}_{max}$ with the corresponding received signal of $\vec y_{max,2}^n$.
Finally,  we denote by $\text{Rx}_{1,1}$  the first $\bar m - n_{max} $ antennas on $\text{Rx}_{1}$ with the corresponding received signal of $\vec y_{1,1}^n$.

We first present the tools that are used to prove the converse for this case.
The first lemma, Least Alignment Lemma, implies that once the transmitter(s) in a network has no CSIT with respect to a certain receiver,  the least amount of alignment will occur at that receiver, meaning that transmit signals will occupy the maximal signal dimensions at that receiver. 
As a result, for the specific case of having two $n_0$-antenna receivers $\text{Rx}_1, \text{Rx}_{2}$, if the transmitter does not have access to any CSIT with respect to $\text{Rx}_{2}$, then the received signal dimension at $\text{Rx}_{2}$ will be greater than $\text{Rx}_{1}$. 

\begin{lemma}\label{lemmamain}
{\bf (Least Alignment Lemma)} Consider two receivers $\Text{Rx}_1,\Text{Rx}_2$ with $n_0$ antennas, where $\text{Rx}_2$ supplies no CSIT.
Then, for a given $n\in \mathbb N$ and  any encoding strategy $ f^{(n)} $ as defined in Definition \ref{coding},
\begin{align*}
h(\vec y_1^n | \mathcal G^n) \leq  h(\vec y_2^n | \mathcal G^n) + n\times o(\log p).
\end{align*}
\end{lemma}

\begin{remark}
Lemma \ref{lemmamain} holds irrespective of the type of CSIT supplied by $\text{Rx}_1$.
Furthermore, Lemma \ref{lemmamain} holds for arbitrary number of transmitters with arbitrary number of transmit antennas; this can be shown via following the same steps as in the  proof  presented in Appendix \ref{LALproof}.
Therefore, Least Alignment Lemma is a general inequality that can be applied to lower bound the received signal dimension at any receiver which supplies no CSIT.
In fact, Least Alignment Lemma  relates the received signal dimension at a receiver which supplies no CSIT to the dimension at other receivers with the same number of antennas, and as a result, proves to be an important tool in analyzing networks with heterogeneous CSIT, where some receivers supply no CSIT, while others supply some form of CSIT to the transmitter(s).
\end{remark}

\cite{OursISIT2014} provided a proof of the lemma for single-antenna receivers, which was limited to networks whose transmitter(s) were only able to employ linear encoding schemes.\footnote{Extension of the proof to networks with multiple antenna receivers  appeared in \cite{OursGlobecom}.} 
However, Davoodi and Jafar \cite{Davoodi} provided the first proof of the inequality in Lemma \ref{lemmamain} (for single-antenna receivers) under general encoding schemes.
Their proof was based on a novel analysis of the Aligned Image Sets at receivers which supply  imperfect CSIT.
The proof was used to settle important conjectures regarding networks with imperfect CSIT \cite{Davoodi}.
Proof of Lemma \ref{lemmamain}, presented in Appendix \ref{LALproof},  extends the result to the MIMO setting.

The following lemma relates the received signal dimensions at 2 receivers supplying no CSIT.

\begin{lemma} \label{ERI nCSIT}
Consider receivers $\Text{Rx}_{1}, \Text{Rx}_{2}$ which supply no CSIT, with $n_{1}, n_2$ antennas, where $n_1 \geq n_{2}$.
Then, for a given $n\in \mathbb N$ and any encoding strategy $f^{(n)} $ as defined in Definition \ref{coding},
\begin{align*}
\frac{h(\vec y_1^n | \mathcal G^n)}{min(m,n_1)} \leq  \frac{h(\vec y_{2}^n | \mathcal G^n)}{min(m,n_{2})} + n\times o(\log p).
\end{align*}
\end{lemma}
Proof of Lemma \ref{ERI nCSIT} follows from channel symmetry, and can be found in prior works, including  \cite{varanasiICNCSIT}.

\begin{proposition} \label{propLAL}
If $m> max(n_1,n_{max})$, then,
\begin{equation*}
h(\vec y_1^n | \mathcal G^n) \stackrel{}{\leq}  \frac{n_1}{\bar n}h(\vec y_{max,1}^n | \mathcal G^n)+ n\times o(\log p). 
\end{equation*}
\end{proposition}
Proof of Proposition \ref{propLAL} follows from Lemma \ref{lemmamain} and Lemma \ref{ERI nCSIT},  and is postponed to Appendix \ref{propLALproof}.

Finally, the following lemma, provides a lower bound on the dimension of joint received signals at a collection of receivers, where some receivers  supply no CSIT.
\begin{lemma} \label{Claimlemma}
Consider receivers $\Text{Rx}_{1}, \Text{Rx}_{2}, \Text{Rx}_{3}$  with $n_{1}, n_2, n_3$ antennas, where $n_1,n_2,n_3>0$, and  $m \geq n_1+  n_{2}+n_3$.
 Further, suppose that $\Text{Rx}_{2}, \Text{Rx}_{3}$ supply no CSIT.
Then, for a given $n\in \mathbb N$ and any encoding strategy $f^{(n)} $ as defined in Definition \ref{coding},
\begin{align*}
\frac{h(\vec y_1^n | \vec y_2^n ,\vec y_3^n, \mathcal G^n)}{n_1} \leq  \frac{h(\vec y_{2}^n | \vec y_3^n, \mathcal G^n)}{n_{2}} + n \times o(\log p).
\end{align*}
\end{lemma}
Lemma \ref{Claimlemma} is proved in Appendix \ref{Claimlemmaproof}. 

\begin{proposition} \label{propclaim}
If $n_1 < n_{max}< m$, for a given $n\in \mathbb N$ and any encoding strategy $f^{(n)} $ as defined in Definition \ref{coding},
\begin{align*}
\frac{h(\vec y_{1,1}^n | \vec y_{max}^n , \mathcal G^n)}{\bar m-n_{max}} \leq  \frac{h(\vec y_{max,2}^n | \vec y_{max,1}^n, \mathcal G^n)}{n_{max}- \bar n} + n\times o(\log p).
\end{align*}
\end{proposition}
Proof of Proposition \ref{propclaim} follows immediately from substituting $(\vec y_1^n , \vec y_2^n ,\vec y_3^n)$ by 
$(\vec y_{1,1}^n ,  \vec y_{max,2}^n , \vec y_{max,1}^n)$ in the statement of Lemma \ref{Claimlemma}.

We now prove the converse for the case where $m> max(n_1,n_{max})$. 
Throughout the proof, we use the notation $h(\emptyset | \mathcal G^n) = 0$ for convenience.
Suppose $d$ secure degrees of freedom is achievable.
By Fano's inequality,
\begin{align}
n(R-\epsilon_n) &\stackrel{}{\leq}  I(W;\vec y_1^n  | \mathcal G^n)
\stackrel{}{\leq}   h(\vec y_1^n   | \mathcal G^n)- h(\vec y_1^n  | W,\mathcal G^n)\nonumber \\
& \substack{\Text{(Lemma \ref{ERI DCSIT})} \\  \leq}  h(\vec y_{1}^n | \mathcal G^n)  - \frac{n_1}{\bar m} h(\vec y_{max}^n  | W,\mathcal G^n) \nonumber\\
& =  h(\vec y_{1}^n | \mathcal G^n)+  \frac{n_1}{\bar m} I(W;\vec y_{max}^n  |\mathcal G^n)  - \frac{n_1}{\bar m} h(\vec y_{max}^n  | \mathcal G^n) \nonumber\\
& \stackrel{(\ref{condition3})}{\leq}  h(\vec y_{1}^n | \mathcal G^n)  - \frac{n_1}{\bar m} h(\vec y_{max}^n  | \mathcal G^n) + n.o(\log p) \nonumber\\
& \stackrel{\text{(Proposition \ref{propLAL})}}{\leq}  \frac{n_1}{\bar n}h(\vec y_{max,1}^n | \mathcal G^n)  - \frac{n_1}{\bar m} h(\vec y_{max}^n  | \mathcal G^n) + n.o(\log p) \nonumber\\
& =  n_1(\frac{\bar m - \bar n}{ \bar m \bar n}) h(\vec y_{max,1}^n | \mathcal G^n) - \frac{n_1}{\bar m} h(\vec y_{max,2}^n |\vec y_{max,1}^n, \mathcal G^n) + n.o(\log p) \nonumber\\
& \leq  n_1(\frac{\bar m - \bar n}{ \bar m })n \log p - \frac{n_1}{\bar m} h(\vec y_{max,2}^n |\vec y_{max,1}^n, \mathcal G^n) + n.o(\log p). \label{mainpartcase2}
\end{align}
We now study the two cases $n_{max}\leq n_{1}$, and $n_{max}>n_1$.
If $n_{max}\leq n_{1}$, then $\frac{n_1}{\bar m} h(\vec y_{max,2}^n |\vec y_{max,1}^n, \mathcal G^n)$, which is the second term on the RHS of (\ref{mainpartcase2}) will equal zero since $\text{Rx}_{max,2}$ is an empty set of antennas when $n_{max}\leq n_{1}$. As a result, since $\bar n=n_{max}$,
\begin{align*}
n(R-\epsilon_n) \stackrel{}{\leq} n_1(\frac{\bar m -  n_{max}}{ \bar m -n_{max}+\bar n})n \log p + n.o(\log p),
\end{align*}
where  by dividing both sides of the  inequality by $n\log p$ and taking the limit $n\to\infty ,p\to \infty$, the converse proof is obtained for the case where $n_{max}\leq n_{1}$.

We now consider the case where $n_{max}>n_1$.
For  this case we  derive a second bound on the secure rate, and then merge it with (\ref{mainpartcase2}) to obtain the converse proof. Again, by Fano's inequality we obtain
\begin{align}
n(R-\epsilon_n) &\stackrel{}{\leq}  I(W;\vec y_1^n  | \mathcal G^n)  \stackrel{}{\leq}  I(W;\vec y_1^n, \vec y_{max}^n  | \mathcal G^n)
  \stackrel{}{\leq}   h(\vec y_1^n, \vec y_{max}^n   | \mathcal G^n)- h(\vec y_{max}^n  | W,\mathcal G^n) \nonumber \\
  &\stackrel{(\ref{condition3})}{=}   h(\vec y_1^n, \vec y_{max}^n   | \mathcal G^n)- h(\vec y_{max}^n  | \mathcal G^n) +n.o(\log p)\nonumber \\
 & \stackrel{(a)}{=}   h(\vec y_{1,1}^n, \vec y_{max}^n   | \mathcal G^n)- h(\vec y_{max}^n  | \mathcal G^n) +n.o(\log p)  =   h(\vec y_{1,1}^n | \vec y_{max}^n  , \mathcal G^n) +n.o(\log p)\nonumber \\
& \stackrel{\text{(Proposition \ref{propclaim})}}{\leq}  (\frac{\bar m - n_{max} }{n_{max}- \bar n })  h(\vec y_{max,2}^n | \vec y_{max,1}^n  , \mathcal G^n) +n.o(\log p), \label{mainpartcase2second} 
\end{align}
where (a) holds since either $m>n_1+n_{max}$, in which case the equality is obvious as $\vec y_{1,1}^n=\vec y_{1}^n$, or $m\leq n_1+n_{max}$, in which case 
given $(\vec y_{1,1}^n, \vec y_{max}^n   ,\mathcal G^n)$,  one can reconstruct the transmit signals within noise level; and as a result, $h(\vec y_{1}^n  | \vec y_{1,1}^n, \vec y_{max}^n   ,\mathcal G^n)= n.o(\log p)$.

We now linearly combine the two inequalities (\ref{mainpartcase2}), (\ref{mainpartcase2second}) and use the fact that $\bar n = n_1$ in this case.
By multiplying both sides of (\ref{mainpartcase2}) by $\frac{\bar m (\bar m-n_{max})}{(\bar m - \bar n )(\bar m-n_{max}+\bar n)}$, and multiplying both sides of (\ref{mainpartcase2second}) by $\frac{ n_1 (n_{max} - \bar n )}{(\bar m - \bar n )(\bar m-n_{max}+\bar n)}$, and then adding the two inequalities together, we obtain the following inequality by considering the assumption that $\bar n = n_1$:
\begin{equation}
n(R-\epsilon_n) \stackrel{}{\leq}  \frac{n_1 (\bar m - n_{max})}{\bar m-n_{max}+\bar n} n \log p +n.o(\log p),\nonumber
\end{equation}
where  by dividing both sides of the  inequality by $n\log p$ and taking the limit $n\to\infty,p\to \infty$, the converse proof is obtained for the case where $n_{max}>n_1$. 
Hence, the proof of converse is complete.

\section{Conclusion} \label{conclusion}
In this paper we  considered the wiretap channel consisting of a legitimate receiver and arbitrary number of eavesdroppers, with delayed CSIT supplied by the legitimate receiver to the transmitter and no eavesdroppers CSIT.
All nodes in the network are equipped with arbitrary number of antennas, hence the name blind MIMOME wiretap channel with delayed CSIT.
We completely characterized the secure Degrees of Freedom (SDoF) for all antenna configurations.

In order to improve the existing achievable scheme we proposed  a two-phase scheme 
where in the first phase artificial noise symbols are transmitted; and in the second phase some of the received noise equations in the first phase are retransmitted together with information symbols in order to drown the information symbols in noise at the eavesdroppers, while the legitimate receiver can decode all the information symbols. 
The signaling makes use of delayed CSIT supplied by the legitimate receiver, and is accomplished in such a way that allows for better artificial noise alignment at the eavesdroppers.
The converse proof is based on several inequalities useful for lower bounding the received signal dimension at receivers with delayed CSIT, or no CSIT, or joint received signal dimension at collections of receivers where some supply no CSIT.

An interesting  future direction is to study the impact of cooperative jamming on the achievable SDoF of blind MIMOME wiretap channel with delayed CSIT, where a jammer is basically a transmitter that does not necessarily have access to the confidential message, but can help jam the confidential message at the eavesdropper(s), hence increase the achievable SDoF.
An initial step towards this direction has been taken in \cite{OursISIT2014}, which studies Blind SISO wiretap channel with cooperative jammer and delayed CSIT.
Another potential follow-up direction to our work in this paper is to consider noisy and delayed CSIT supplied by the legitimate receiver, rather than perfect and delayed CSIT, and study how this impacts the achievable SDoF.

\begin{appendices}

\section{proof of Least Alignment Lemma (Lemma \ref{lemmamain})} \label{LALproof}
We first restate Lemma \ref{lemmamain} here.

\begin{lemmarep}
{\bf Lemma \ref{lemmamain} (Least Alignment Lemma)} 
Consider two receivers $\Text{Rx}_1,\Text{Rx}_2$ with $n_0$ antennas, where $\text{Rx}_2$ supplies no CSIT.
Then, for a given $n\in \mathbb N$ and any encoding strategy $ f^{(n)} $ as defined in Definition \ref{coding},
\begin{align*}
h(\vec y_1^n | \mathcal G^n) \leq  h(\vec y_2^n | \mathcal G^n) + n.o(\log p).
\end{align*}
\end{lemmarep}
We prove a stronger version of Lemma \ref{lemmamain}. 
More specifically, we prove that 
if $\text{Tx}$ only knows a probability distribution for values of the channels to $\text{Rx}_2$, and we denote the maximum value of such distribution by $f_{max}(p)$,
 then
\begin{align*}
h(\vec y_1^n | \mathcal G^n) \leq  h(\vec y_2^n | \mathcal G^n) + nn_0\log(f_{max}(p))+ n.o(\log p).
\end{align*}
In order to prove the above inequality, we use the approach  in \cite{Davoodi} to generalize that result  to the MIMO case.
In particular, we first  transform the network via  a deterministic channel model.

\subsection{Deterministic Channel Model}
To prove the lemma, we first discretize the channel to avoid dealing with the impact of additive Gaussian noise.
This leads to a deterministic channel model described as follows.
For $j=1,2,$ let
\begin{align}
\vec{\bar x}(t) &= \left [  \begin{tabular}{ccc}
  $\bar x_{1}(t)$ \\  \vdots \\$\bar x_{m}(t)$  \end{tabular} \right ],  \qquad 
\vec{\bar y}_j(t) &= \left [  \begin{tabular}{ccc}
  $\bar y_{j,\{1\}}(t)$ \\  \vdots \\$\bar y_{j,\{n_0\}}(t)$  \end{tabular} \right ] , \qquad
G_{j}(t) = \left [  \begin{tabular}{ccc}
  $g_{j,\{1,1\}}(t)$  & $\ldots$ & $g_{j,\{1,m\}}(t)$  \\  \vdots \\$g_{j,\{n_0,1\}}(t)$  & $\ldots$ & $g_{j,\{n_0,m\}}(t)$ \end{tabular} \right ].
\end{align} 

The channel input at time $t$ is denoted by $\vec {\bar x}(t)$, where $\vec {\bar x}(t)\in \{ 0,1 , \ldots , \lceil \sqrt{p} \rceil  \}^m$ .
The channel outputs are defined as
\begin{align}
\vec{\bar y}_j(t) &= \left [  \begin{tabular}{ccc}
  $\bar y_{j,\{1\}}(t)$ \\  \vdots \\$\bar y_{j,\{n_0\}}(t)$  \end{tabular} \right ]  = 
\left [  \begin{tabular}{ccc}
  $\sum_{i=1}^{m} \lfloor  g_{j,\{1,i\}}(t)  \bar x_i(t) \rfloor$ \\  \vdots \\$\sum_{i=1}^{m} \lfloor  g_{j,\{n_0 ,i\}}(t)  \bar x_i(t) \rfloor$  \end{tabular} \right ]  .
\end{align}

\begin{lemma} \label{deterministic}
\begin{align}
\max \lim_{p\to\infty} \frac{h(\vec y_1^n | \mathcal G^n) -  h(\vec y_2^n | \mathcal G^n)}{\log p} \leq 
\max \lim_{p\to\infty} \frac{H(\vec{\bar y}_1^n | \mathcal G^n) -  H(\vec{\bar y}_2^n | \mathcal G^n)}{\log p},
\end{align}
where the maximum on the left hand side  is taken over all possible encoding strategies as defined in Definition \ref{coding}; and the maximum on the right hand side is taken over all possible encoding schemes for the deterministic channel.
\end{lemma}

Proof of Lemma \ref{deterministic} follows from similar arguments used to prove that DoF of a network under deterministic channel model is an upper bound on the actual DoF.
The proof can be found in prior works, including \cite{ BreslerDet, Davoodi}, and hence is omitted for brevity.
Lemma \ref{deterministic} suggests that for  proving Lemma \ref{lemmamain}, it is sufficient to prove that under deterministic channel model, 
\begin{align}
H(\vec{\bar y}_1^n | \mathcal G^n) -  H(\vec{\bar y}_2^n | \mathcal G^n) \leq n.o(\log p). \label{altdet}
\end{align}
As a result, our objective henceforth will be to  prove (\ref{altdet}).

\subsection{Imposing Functional Dependence}
We will show in this section that by imposing functional dependence of $\vec{\bar x}^n$ on $(\vec{\bar y}_1^n, G_1^n)$, we obtain an upper bound on $H(\vec{\bar y}_1^n | \mathcal G^n) -  H(\vec{\bar y}_2^n | \mathcal G^n) $.
Define $\mathcal L$ as the mapping from  $(\vec{\bar y}_1^n, G_1^n)$ to $\vec{\bar x}^n$:
\begin{align}
\vec{\bar x}^n = \mathcal L(\vec{\bar y}_1^n,  G_1^n).
\end{align}
 This mapping is in general stochastic, and therefore, $\mathcal L$ is a random variable.
Hence, by conditioning on $\mathcal L$ we obtain
\begin{align}
 H(\vec{\bar y}_2^n | \mathcal G^n) &\geq   H(\vec{\bar y}_2^n | \mathcal G^n, \mathcal L)\geq   \min_{L} H(\vec{\bar y}_2^n | \mathcal G^n,\mathcal L= L)  =   H(\vec{\bar y}_2^n | \mathcal G^n,\mathcal L= L_0),  \label{Lcond}
\end{align}
where\footnote{In cases where $\arg\min_{L}  H(\vec{\bar y}_2^n | \mathcal G^n,\mathcal L= L)$ is not unique, we choose $L_0$ to be a deterministic mapping that minimizes $H(\vec{\bar y}_2^n | \mathcal G^n,\mathcal L= L)$. 
}
$ 
L_0 \triangleq \arg\min_{L}  H(\vec{\bar y}_2^n | \mathcal G^n,\mathcal L= L)
$
is a deterministic map.
Note that the choice of map does not impact $(\vec{\bar y}_1^n, \mathcal G^n)$.
Hence, using (\ref{Lcond}) we obtain
\begin{align}
H(\vec{\bar y}_1^n | \mathcal G^n) -  H(\vec{\bar y}_2^n | \mathcal G^n) \leq    H(\vec{\bar y}_1^n | \mathcal G^n,\mathcal L= L_0)  -   H(\vec{\bar y}_2^n | \mathcal G^n,\mathcal L= L_0). \label{funcdep}
\end{align}
Thus, henceforth, we will upper bound (\ref{funcdep}) in order to complete the proof of Lemma \ref{lemmamain}; and we will assume that 
\begin{align}
\vec{\bar x}^n = L_0(\vec{\bar y}_1^n, G_1^n).
\end{align}
Note that the above equation suggests that $\vec{\bar y}_2^n$ is fully specified by $(\vec{\bar y}_1^n, \mathcal G^n)$; 
hence, we use the following notation for  simplicity:
\begin{align}
\vec{\bar y}_2^n(\vec{\bar y}_1^n, \mathcal G^n) \triangleq \left [  \begin{tabular}{ccc}
  $\sum_{i=1}^{m} \lfloor  g_{2,\{1,i\}}(t)   L_0(\vec{\bar y}_1^n,  G_1^n)_i(t) \rfloor$ \\  \vdots \\$\sum_{i=1}^{m} \lfloor  g_{2,\{n_0 ,i\}}(t)  L_0(\vec{\bar y}_1^n,  G_1^n)_i(t) \rfloor$  \end{tabular} \right ],\label{defofy2}
\end{align}
where $L_0(\vec{\bar y}_1^n, G_1^n)= \vec {\bar x}^n$, and $L_0(\vec{\bar y}_1^n, G_1^n)_i(t)= \bar x_i(t)$, which is the i-th element of the vector $\vec{\bar x}(t)$.

\subsection{Upper Bounding $H(\vec{\bar y}_1^n | \mathcal G^n) -  H(\vec{\bar y}_2^n | \mathcal G^n)$ via Aligned Image Sets} \label{ubLALdiff}
Note that our goal is to upper bound $H(\vec{\bar y}_1^n | \mathcal G^n) -  H(\vec{\bar y}_2^n | \mathcal G^n)$.
This means that we will try to upper bound the difference between received signal dimensions at $\text{Rx}_1, \text{Rx}_2$. 
This difference grows when more codewords that are received at $\text{Rx}_1$ as different received codewords align perfectly at $\text{Rx}_2$; because in such case, the dimension of received signal at $\text{Rx}_2$ will decrease, leading to an increase in the difference of received signal dimensions at the two receivers.
Hence, we will focus on aligned images sets, which are sets of distinct codewords received at $\text{Rx}_1$ that are aligned at $\text{Rx}_2$.
More specifically, 
for a pair  $(\vec{\bar v}^n, \mathcal G^n)$ of received codeword at $\text{Rx}_2$, $\vec{\bar v}^n$, and channel coefficients, $\mathcal G^n$, we define the corresponding aligned image set as the set of all received signals at $\text{Rx}_1$ which have the same image  at $\text{Rx}_2$ as $\vec{\bar v}^n$.
\begin{definition}
{\bf (Aligned Image Set) }
$S_{\vec{\bar v}^n}( \mathcal G^n) \triangleq \Bigg \{  \vec{\bar y}_1^n   \quad  |  \quad     
\vec{\bar y}_2^n(\vec{\bar y}_1^n, \mathcal G^n)
= \vec{\bar v}^n
, \quad  t=1,\ldots , n  \Bigg \}.$
\end{definition}

We now  upper bound $H(\vec{\bar y}_1^n | \mathcal G^n) -  H(\vec{\bar y}_2^n | \mathcal G^n)$ via analyzing the cardinality of aligned image sets.
\begin{align}
H(\vec{\bar y}_1^n | \mathcal G^n) -  H(\vec{\bar y}_2^n | \mathcal G^n)
& \leq   H(\vec{\bar y}_1^n, \vec{\bar y}_2^n  | \mathcal G^n, L_0) -  H(\vec{\bar y}_2^n | \mathcal G^n, L_0)=    H(\vec{\bar y}_1^n | \vec{\bar y}_2^n  , \mathcal G^n, L_0)   \nonumber\\
&\stackrel{(a)}{=}   H(\vec{\bar y}_1^n | \quad |S_{\vec{\bar y}_2^n}(\mathcal G^n)|,\vec{\bar y}_2^n, \mathcal G^n, L_0)        \nonumber\\
&\stackrel{(b)}{\leq}   E \log |S_{\vec{\bar y}_2^n}(\mathcal G^n)|        \nonumber\\
&\stackrel{(c)}{\leq}   \log E|S_{\vec{\bar y}_2^n}(\mathcal G^n)|,       \label{ExpObj}
\end{align}
where
(a) holds since $( \vec{\bar y}_2^n, \mathcal G^n, L_0)$ completely determines the aligned image set $ S_{\vec{\bar y}_2^n}(\mathcal G^n)$;
(b) holds since the entropy of $\vec{\bar y}_1^n$ is maximized when it has a uniform distribution over all of its possible values, which are determined by $S_{\vec{\bar y}_2^n}(\mathcal G^n)$;
and (c) holds due to Jensen's inequality.

Hence, we will focus on upper bounding  $ E|S_{\vec{\bar y}_2^n}(\mathcal G^n)|$.
To this aim, for a  given $(\vec{\bar y}_1^n,\vec{\bar v}^n)$ we first analyze  $\Pr(\vec{\bar y}_1^n \in S_{\vec{\bar v}^n}(\mathcal G^n))$, which is the probability that the received image of a certain codeword at $\text{Rx}_1$ has the same image at $\text{Rx}_2$ as $\vec{\bar v}^n$.

\subsection{Bounding the probability of Image Alignment}
In this section we will provide an upper bound on $\Pr(\vec{\bar y}_1^n \in S_{\vec{\bar v}^n}(\mathcal G^n))$.
To this aim, we analyze $\Pr(\vec{\bar y}_1^n \in S_{\vec{\bar v}^n}(\mathcal G^n)| G_1^n)$.
Let us fix $\vec{\bar y}_1^n , \vec{\bar v}^n, G_1^n$. 
Note that given $(\vec{\bar y}_1^n, G_1^n)$, $\vec{\bar x}^n$ is determined.
Consider the event where $\vec{\bar y}_1^n \in S_{\vec{\bar v}^n}(\mathcal G^n)$. 
This event is equivalent to $\vec{\bar y}_2^n (\vec{\bar y}_1^n ,\mathcal G^n)= \vec{\bar v}^n$, which in turn by (\ref{defofy2}) is equivalent to the following:
\begin{align}
\forall t=1,\ldots, n, \qquad \left [  \begin{tabular}{ccc}
  $\sum_{i=1}^{m} \lfloor  g_{2,\{1,i\}}(t)   L_0(\vec{\bar y}_1^n,  G_1^n)_i(t) \rfloor$ \\  \vdots \\$\sum_{i=1}^{m} \lfloor  g_{2,\{n_0 ,i\}}(t)  L_0(\vec{\bar y}_1^n, G_1^n)_i(t) \rfloor$  \end{tabular} \right ] = \vec{\bar v}(t) 
= \left [  \begin{tabular}{ccc}
  $\bar v_1(t)$ \\  \vdots \\ $\bar v_{n_0}(t)$  \end{tabular} \right ], \label{event}
\end{align}
or equivalently,
\begin{align}
\forall t=1,\ldots, n,  j=1,\ldots, n_0 , \qquad 
\sum_{i=1}^{m} \lfloor  g_{2,\{j,i\}}(t)   \bar x_i(t) \rfloor  = \bar v_j(t).\label{event2}
\end{align}
Let $i^*(t) = \arg \max_{i}\bar x_i(t) $.
As a result, the above event can be re-written as follows:
\begin{align}
&\forall t=1,\ldots, n,  j=1,\ldots, n_0, \nonumber\\
&\bar v_j(t) -  \sum_{i=1, i\ne i^*(t)}^{m} \lfloor  g_{2,\{j,i\}}(t)   \bar x_i(t) \rfloor \leq  g_{2,\{j,i^*(t)\}}(t)  \bar x_{i^*(t)}(t) \leq  \bar v_j(t) -  \sum_{i=1, i\ne i^*(t)}^{m} \lfloor  g_{2,\{j,i\}}(t)   \bar x_i(t) \rfloor + 1.\nonumber
\end{align}
Hence, for every $t$, if $\bar x_{i^*(t)}(t) \ne  0$, then for the event (\ref{event2}) to occur  it is necessary that $ g_{2,\{j,i^*(t) \}}(t)$ fall in an interval of length   $\frac{1}{\bar x_{i^*(t)}(t) }$ for $t=1,\ldots, n,  j=1,\ldots, n$.
Also, note that if $\vec{\bar x}(t) = \vec 0$ (which means $\bar x_{i^*(t)}(t) =  0$), then $\vec{\bar y}_1(t) = \vec{\bar y}_2(t)  =  \vec 0$.
Hence, for all $t=1,\ldots, n,$ where $ \vec{\bar y}_1(t)  \ne \vec 0$, the probability of occurrence of (\ref{event2}) for $t=1,\ldots, n,  j=1,\ldots, n_0,$ is at most $f_{max}(p) (\frac{1}{\bar x_{i^*(t)}(t) })$.
We now further upper bound this quantity.
Note that since
$\bar y_{1,\{j\} }(t) = \sum_{i=1}^{m} \lfloor  g_{1,\{j,i\}}(t)   \bar x_i(t) \rfloor$, 
\begin{align}
|\bar y_{1,\{j\} }(t)| \leq    \bar x_{i^*(t)}(t)  \sum_{i=1}^{m}  |g_{1,\{j,i\}}(t)|   + m.\nonumber
\end{align}
By re-writing the above inequality, when $|\bar y_{1,\{j\} }(t)| > m$,
\begin{align}
\frac{1}{ \bar x_{i^*(t)}(t)} \leq  \frac{ \sum_{i=1}^{m}  |g_{1,\{j,i\}}(t)|   }{ |\bar y_{1,\{j\} }(t)| - m}.\nonumber
\end{align}

Hence, we have the following upper bound on the probability of occurrence of $\vec{\bar y}_1^n \in S_{\vec{\bar v}^n}(\mathcal G^n)$:
\begin{align}
\Pr(\vec{\bar y}_1^n \in S_{\vec{\bar v}^n}(\mathcal G^n)| G_1^n)
 &\leq \prod_{\substack{t:\\ \vec{\bar y}_1(t)  \ne \vec 0}} \prod_{\substack{j:\\ |\bar y_{1,\{j\} }(t)| > m}}   \frac{ f_{max}(p)( \sum_{i=1}^{m}  |g_{1,\{j,i\}}(t)| )  }{ |\bar y_{1,\{j\} }(t)| - m}\nonumber\\
&\leq  \left ( \prod_{t=1}^{n} \prod_{j=1}^{n}  \max (1, f_{max}(p)( \sum_{i=1}^{m}  |g_{1,\{j,i\}}(t)| )) \right )
  \prod_{t=1}^{n} \prod_{j=1}^{n}   \frac{ 1  }{ \max(1, |\bar y_{1,\{j\} }(t)| - m) }\nonumber \\
& \leq  \max (1, f_{max}(p) md_{max})^{nn_0}
 \prod_{t=1}^{n} \prod_{j=1}^{n}   \frac{ 1  }{ \max(1, |\bar y_{1,\{j\} }(t)| - m) }. \label{probbound}
\end{align}
We use the above bound on $\Pr(\vec{\bar y}_1^n \in S_{\vec{\bar v}^n}(\mathcal G^n)| G_1^n)$ to further upper bound (\ref{ExpObj}).

\subsection{Bounding the Average  Size of Aligned Image Sets} \label{avesize}
For a given $\vec{\bar v}^n$,
\begin{align}
E [ S_{\vec{\bar v}^n}(\mathcal G^n)] &=  E[ E [ S_{\vec{\bar v}^n}(\mathcal G^n)|G_1^n]] 
= E[ \sum_{\vec{\bar y}_1^n}^{} \Pr  (\vec{\bar y}_1^n \in S_{\vec{\bar v}^n}(\mathcal G^n)| G_1^n) ]\nonumber\\
& \stackrel{(\ref{probbound})}{\leq}  \max (1, f_{max}(p) m d_{max})^{nn_0}   \sum_{\vec{\bar y}_1^n}^{}
 \prod_{t=1}^{n} \prod_{j=1}^{n}   \frac{ 1  }{ \max(1, |\bar y_{1,\{j\} }(t)| - m) }  \nonumber\\
& \stackrel{(a)}{= } \max (1, f_{max}(p)   m d_{max})^{nn_0}  
 \prod_{t=1}^{n} \prod_{j=1}^{n}   \sum_{\bar y_{1,\{j\} }(t)}^{}  \frac{ 1  }{ \max(1, |\bar y_{1,\{j\} }(t)| - m) } \nonumber 
\end{align}
\begin{align}
 &\stackrel{(b)}{\leq } \max (1, f_{max}(p)  md_{max})^{nn_0}  
 \prod_{t=1}^{n} \prod_{j=1}^{n}  \{ \sum_{\substack{\bar y_{1,\{j\} }(t):\\ |\bar y_{1,\{j\} }(t)| \leq m }}^{} 1 + \sum_{\substack{\bar y_{1,\{j\} }(t):\\ m<|\bar y_{1,\{j\} }(t) |\leq q }}^{} \frac{ 1  }{  |\bar y_{1,\{j\} }(t)| - m } \}  \nonumber \\
 &\stackrel{}{\leq } \max (1, f_{max}(p)md_{max})^{nn_0}  
 \prod_{t=1}^{n} \prod_{j=1}^{n}  \{ \log q + o(\log q) \}  \nonumber\\
 &\stackrel{}{\leq } \max (1, f_{max}(p) m d_{max})^{n n_0}  (\log q)^{nn_0}+ o(\log q) \nonumber \\
 &\stackrel{}{\leq } \max (1, f_{max}(p)md_{max})^{n n_0}  (\log \sqrt{p})^{n n_0}+ o(\log p),\label{avesizeineq}
\end{align}
where (a) follows from interchanging sum and product;
and (b) follows from the definition $q\triangleq md_{max}\sqrt{p} + m$, and noting that $|\bar y_{1,\{j\} }(t) |\leq q$.
Hence, by (\ref{ExpObj}) and (\ref{avesizeineq}), we obtain
\begin{align}
H(\vec{\bar y}_1^n | \mathcal G^n) -  H(\vec{\bar y}_2^n | \mathcal G^n) 
&\stackrel{}{\leq} \log\left( \max (1, f_{max}(p) m d_{max})^{nn_0}  (\log \sqrt{p})^{nn_0}+ o(\log p) \right) \nonumber \\
&= nn_0\log(f_{max}(p)) + o(\log p). \label{endofLALproof}
\end{align}
Therefore,  the proof of Lemma \ref{lemmamain} is complete.

\begin{remark}
Note that (\ref{endofLALproof}) in fact proves a stronger statement than  Lemma \ref{lemmamain}.
In particular, for any CSIT quality supplied by $\text{Rx}_2$ as a function of power $f_{max}(p)$, (\ref{endofLALproof}) implies that
\begin{align*}
h(\vec y_1^n | \mathcal G^n) \leq  h(\vec y_2^n | \mathcal G^n) + nn_0\log(f_{max}(p)) + n.o(\log p).
\end{align*}
Nevertheless, in order to prove Theorem \ref{maintheorem}, it was sufficient to consider the special case where $f_{max}(p)= o(\log p)$, which is the case in the statement of Lemma \ref{lemmamain}.
\end{remark}

\section{Proof of Proposition \ref{propLAL}} \label{propLALproof}
Let us first consider a hypothetical receiver $\Text{Rx}_{0}$ with $n_1$ antennas for which there is no CSIT available to the transmitter.
Hence, by Lemma \ref{lemmamain},
\begin{equation} \label{MIMOLAI}
h(\vec y_1^n | \mathcal G^n) \leq  h(\vec y_0^n | \mathcal G^n) + n.o(\log p).
\end{equation} 
Moreover, since $n_1\geq \bar n$ and there is no CSIT with respect to any of $\Text{Rx}_{0},\Text{Rx}_{max,1}$, using Lemma \ref{ERI nCSIT},
\begin{equation} \label{MIMOLemma1}
\frac{  h(\vec y_0^n | \mathcal G^n)  }{n_1} \stackrel{}{\leq}  \frac{ h(\vec y_{max,1}^n | \mathcal G^n) }{\bar n} + n.o(\log p).
\end{equation}
Therefore, by combining the inequalities in (\ref{MIMOLAI})-(\ref{MIMOLemma1}) we get
\begin{equation}\label{RRIn1geqn2}
h(\vec y_1^n | \mathcal G^n) \stackrel{}{\leq}  \frac{n_1}{\bar n}h(\vec y_{max,1}^n | \mathcal G^n) + n.o(\log p),\nonumber
\end{equation}
which completes the proof of Proposition \ref{propLAL}.

\section{Proof of Lemma \ref{Claimlemma}}   \label{Claimlemmaproof}
We first state an extension of Lemma \ref{ERI nCSIT}, which  is useful in proving Lemma \ref{Claimlemma}, and can be proved using the same proof steps as in proof of Lemma \ref{ERI nCSIT}.

\begin{lemma} \label{ERI nCSITExt}
Consider receivers $\Text{Rx}_{1}, \Text{Rx}_{2}, \Text{Rx}_{3}$ which supply no CSIT, with $n_{1}, n_2, n_3$ antennas, where $n_1 \geq n_{2}$.
Then, for a given $n\in \mathbb N$ and any encoding strategy $f^{(n)} $ as defined in Definition \ref{coding},
\begin{align*}
\frac{h(\vec y_1^n | \vec y_3^n, \mathcal G^n)}{min(m,n_1)} \leq  \frac{h(\vec y_{2}^n | \vec y_3^n, \mathcal G^n)}{min(m,n_{2})} + n\times o(\log p).
\end{align*}
\end{lemma}

Consider receivers $\Text{Rx}_{1}, \Text{Rx}_{2}, \Text{Rx}_{3}$  with $n_{1}, n_2, n_3$ antennas, where $n_1,n_2,n_3>0$, and  $m \geq n_1+  n_{2}+n_3$. 
Also, suppose that $\Text{Rx}_{2}, \Text{Rx}_{3}$ supply no CSIT.
Further, let $\Text{Rx}_{0}$ denote a receiver with $n_1$ antennas supplying no CSIT.

\begin{align}
n_2\times h(\vec y_1^n | \vec y_2^n ,\vec y_3^n, \mathcal G^n) & = n_2\times h(\vec y_1^n , \vec y_2^n ,\vec y_3^n| \mathcal G^n) - n_2\times h(\vec y_3^n | \mathcal G^n) - n_2\times h(\vec y_2^n |\vec y_3^n, \mathcal G^n)\nonumber \\
& \stackrel{(a)}{\leq} n_2\times h(\vec y_{0}^n , \vec y_2^n ,\vec y_3^n| \mathcal G^n) - n_2\times h(\vec y_3^n | \mathcal G^n) - n_2\times h(\vec y_2^n |\vec y_3^n, \mathcal G^n)+ n.o(\log p) \nonumber \\
& = n_2\times h(\vec y_{0}^n , \vec y_2^n | \vec y_3^n,  \mathcal G^n) - n_2\times h(\vec y_2^n |\vec y_3^n, \mathcal G^n) + n.o(\log p) \nonumber\\
& \stackrel{\text{Lemma \ref{ERI nCSITExt}}}{\leq}  n_2 \times (\frac{n_1+n_2}{n_2}) \times h(  \vec y_2^n | \vec y_3^n,  \mathcal G^n) - n_2\times h(\vec y_2^n |\vec y_3^n, \mathcal G^n)+ n.o(\log p) \nonumber \\
& \leq  n_1 \times h(\vec y_{2}^n | \vec y_3^n, \mathcal G^n) + n.o(\log p), \label{lastlineclaimlemma}
\end{align}
where (a) holds since the virtual receiver incorporating $(\text{Rx}_{0},\text{Rx}_2,\text{Rx}_3)$, which has $n_1+n_2+n_3$ antennas, supplies no CSIT; and as a result, we can apply Lemma \ref{lemmamain} to lower bound $h(\vec y_{0}^n , \vec y_2^n ,\vec y_3^n| \mathcal G^n)$ by $h(\vec y_{1}^n , \vec y_2^n ,\vec y_3^n| \mathcal G^n)$.
By rearranging both sides of (\ref{lastlineclaimlemma}), proof of Lemma \ref{Claimlemma} will be complete.

\end{appendices}

\bibliographystyle{ieeetr}
\bibliography{bib_delayedCSIT}

\end{document}